\documentclass{aa}
\usepackage{epsfig}
\usepackage{times}
\usepackage{mathptm}
\epsfclipon
\begin{document}

   \thesaurus{12     
              (12.07.1;  
               11.17.3;  
               11.17.4)} 
   \title{First estimate of the time delay in HE~1104--1805
          \thanks{Based on observations collected at the European
                  Southern Observatory, La Silla, Chile}}

   \author{Lutz Wisotzki\inst{1}
           \and
           Olaf Wucknitz\inst{1}
           \and
           Sebastian Lopez\inst{1}
           \and
           Anton Norup S\o{}rensen\inst{2}
          }

   \offprints{L. Wisotzki, lwisotzki@hs.uni-hamburg.de}

   \institute{
              Hamburger Sternwarte, Universit\"at Hamburg, Gojenbergsweg 112, 
              21029 Hamburg, Germany
         \and
              Copenhagen University Observatory, Juliane Maries Vej 30
              DK-2100 Copenhagen, Denmark
             }

   \date{Received; accepted}

   \maketitle

   \begin{abstract}

      We present first results from five years of spectrophotometric
      monitoring of the bright double QSO and gravitational lens
      HE~1104$-$1805. The quasar has varied considerably over this
      time, while the emission line fluxes appear to have
      remained constant. We have constructed monochromatic continuum
      light curves for components A and B, finding that B leads
      the variability. A quantitative analysis with the Pelt method 
      gives a best estimate for the light travel time delay 
      of about 0.73 years, although a value as low as 0.3 cannot
      yet be excluded. We discuss possible models for the QSO-lens 
      configuration and use our measured time delay to predict the
      redshift of the lens, $z_\mathrm{d}$. Finding that most likely
      $z_\mathrm{d}\la 1$, we can rule out the 
      hitherto favoured values of $z_\mathrm{d} = 1.32$ or 1.66.
      A new candidate is an absorption system at $z=0.73$,
      but the lens could also be an elliptical not detected
      in absorption.

      \keywords{Quasars: individual: HE 1104$-$1805 --
                Quasars: general --
                Gravitational lensing
               }
   \end{abstract}

\section{Introduction}

HE~1104$-$1805 (the `Double Hamburger', $z=2.32$) is one of the 
brightest multiply imaged QSOs in the sky: Around discovery in 1993, 
its $B$ band magnitudes were 16.7 and 18.6 for components A and B,
respectively (Wisotzki et al.\ \cite{wisotzki93}).
The gravitational lens nature is now firmly
established, after the deflecting galaxy has been detected
in the line of sight (Courbin et al.\ \cite{courbin98};
Remy et al.\ \cite{remy98}, quoted as R98 in the following),
but the lens redshift is still unknown. The strong
$z=1.66$ damped Ly$\alpha$ system in component A (Smette et al.\ 
\cite{smette*95}) may be an obvious candidate, but the observed
properties do not agree well with this hypothesis 
(R98; Lopez et al.\ \cite{lopez98}).

Because of their brightness and the relatively large image 
separation of $3\farcs 2$, photometry of the two components 
can be obtained under less than optimal circumstances,
especially as HE~1104$-$1805 shows variability 
with considerable amplitude (Wisotzki et al.\ \cite{wisotzki95}).
The system is therefore well-suited for systematic monitoring,
with the ultimate aim to obtain the light travel time delay
and to estimate the Hubble parameter. We have started with a
\emph{spectrophotometric} monitoring in early 1996, 
and including some earlier observations the time span covered 
is now five years.
Here we present first results from this monitoring campaign,
focused specifically on the issue of time delay estimation.
Other aspects of the monitoring, in particular 
a comparison of continuum and emission line properties and the
possible signatures of microlensing, will be 
dealt with in a future paper.

\section{Observations and data reduction}

The spectrophotometric monitoring was conducted at the ESO 3.6\,m
telescope in service mode. Typically
once a month during the visibility period,
up to one hour was dedicated to the programme.
The instrument was EFOSC1 with 512$\times$512 pixels Tektronix 
CCD until June 1997, and EFOSC2 with a 2K$\times$2K Loral/Lesser
chip afterwards. The observations always comprised one or several
acquisition images, followed by a low-resolution spectrum with 
the B300 grism through a $5''$ slit aligned along components A and B.
Exposure time was between 15--30\,min, spectral resolution was
typically $\sim 20$\,\AA , and the spectra covered the wavelength 
range between 3800\,\AA\ and 7000\,\AA . 
The seeing was between $1\farcs 1$ and $2\farcs 0$. 
A few datasets have been
included in this analysis that were obtained before the proper
monitoring started. Some were taken with the same configuration
as described above; in addition we used the NTT data shown already by 
Wisotzki et al.\ (\cite{wisotzki93}) and a set of spectra obtained
with the LDS spectrograph at the Nordic Optical Telescope 
in February 1994.

The CCD frames were reduced in a homogeneous way, largely following
standard procedures. The most critical task was to perform
an unbiased simultaneous extraction of the two components. We have
employed a three-step deblending algorithm, 
consisting of the following tasks: (1) Two Gaussians,
of the same FWHM and with fixed angular separation 
but otherwise unconstrained, 
were fitted simultaneously to each spectral resolution element. 
(2) The variation with wavelength of the resulting parameters FWHM and 
centroid location was fitted by low-order polynomials.
(3) Another double-Gaussian fit was performed, now with only the two
amplitudes as free parameters. 
The algorithm is described in more detail 
by Lopez et al.\ (\cite{lopez98}).
Inspection of the residual maps
revealed no significant deviation from this model. The resulting
one-dimensional spectra were recorded as MIDAS table files, thus
avoiding loss of information due to data rebinning.

Wavelength calibration frames were obtained
from comparison lamp spectra, usually based on Helium/Argon lines.
The observing conditions were in many cases not photometric, and
standard star spectra for flux calibration purposes were obtained in only
a few nights. Therefore only a relative flux calibration could be
attempted. Two of the pre-monitoring observations were made
through a narrow slit, which in one case was not even aligned.
We explain below how these measurements could be incorporated 
into the analysis.

\begin{figure}
\epsfxsize=8.8cm\epsfbox[121 127 356 299]{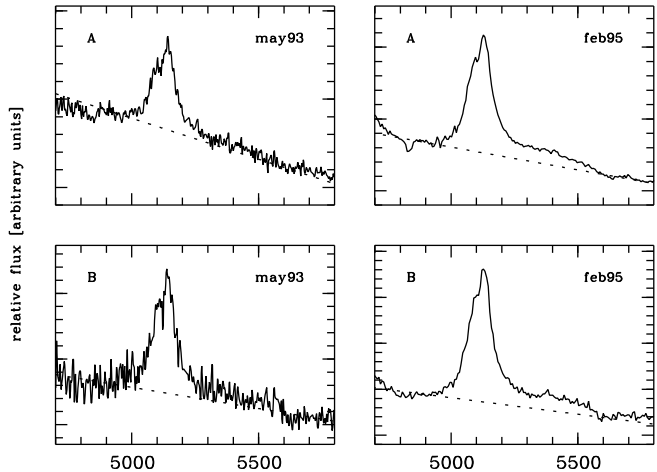}
\caption[]{Example spectra obtained in the monitoring, showing the 
           region around C\,{\small IV} $\lambda$1549
           for two epochs, in components A and B.
           The adopted continua are marked by the dotted lines.}
   \label{fig:spectra}
\end{figure}

\begin{figure}
\epsfxsize=8.8cm\epsfbox[75 114 455 455]{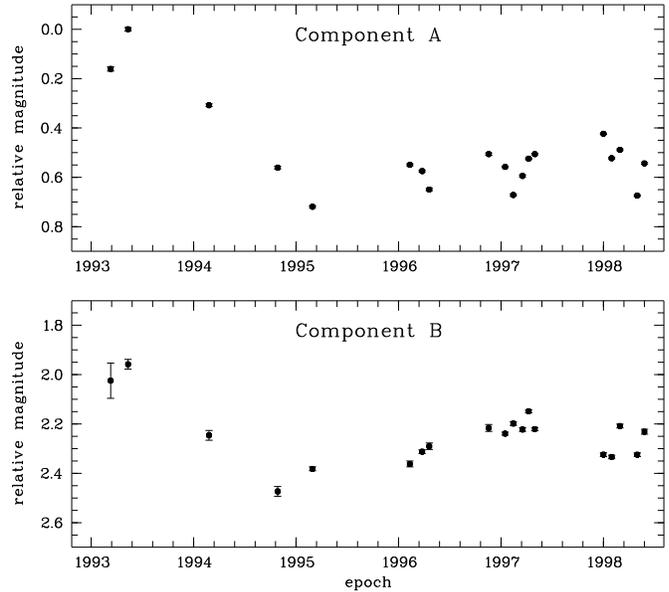}
\caption[]{Monochromatic continuum light curves of both components,
           taken at $\lambda_c = 4910$\,\AA . The zero point is
           arbitrary, but the same for both panels. Error bars
           in the upper panel are generally smaller than the symbol 
           size.
           }
   \label{fig:contlc}
\end{figure}

\section{Continuum light curves}

Altogether, 19 sets of spectra could be secured, of which 14
were obtained in the course of the monitoring. Most spectra are of 
very high quality, some with a continuum S/N ratio exceeding 100
in component A. Examples of the spectra are presented in Fig.\
\ref{fig:spectra}; the range in S/N ratio is bracketed by 
these example data, with most spectra looking rather similar 
to the February 1995 ones.
The high quality of the data enabled us to monitor the
\emph{monochromatic continuum fluxes} of both components,
which has the advantage over the conventional broad-band
magnitudes that it is a brightness measure uncontaminated 
by emission lines and their possibly different
variability patterns.
However, absolute photometric calibration from standard stars
was usually not possible, and we had to design a method 
to compare spectra taken at different epochs.
In the following we motivate and outline this procedure.

After placing the spectra on a relative flux scale, we measured
the fluxes and equivalent widths of all major emission lines
in both components (Ly$\alpha$, Si\,{\small IV} $\lambda$1400, 
C\,{\small IV} $\lambda$1549, and C\,{\small III}] $\lambda$1909).
Local continua were estimated by fitting straight lines to
predefined wavebands known to be largely devoid of emission
and absorption lines (cf.\ examples in Fig.\ \ref{fig:spectra}).
Although a direct comparison of line strengths between different 
epochs is not possible, there is one strong piece of evidence that the lines 
have remained essentially constant over the time span observed:
The flux ratio between the same lines in components A and B, 
independent of the absolute scale, has stayed at a 
consistently constant value of $2.85\pm 0.07$ for all lines. 
This implies either a time delay of much less than a month
(the separation between data points during the periods of
quasi-continuous monitoring), which is highly improbable, 
or simply constancy of the line fluxes as such. 
Adopting the latter hypothesis, we were then able to recalibrate 
the spectra by scaling them to equal
emission line fluxes. As reference we used C\,{\small IV},
a prominent line that is surrounded by clearly
identifiable continuum windows visible also at low spectral
resolution. 

For each pair of spectra we computed a scale factor so that the
C\,{\small IV} flux of component A assumed an arbitrary but constant
value, and applied this factor to both spectra. The average continuum
values in the interval $4880\,\mbox{\AA}<\lambda<4940\,\mbox{\AA}$,
thus at $\lambda_c = 4910$\,\AA , were then determined for A and B.
This estimate of the QSOs brightness is
independent of external flux standards and
of photometric conditions, and we were thus able to incorporate 
also narrow slit observations by the same method.
Note that the relative flux calibration based on standard
stars was not even strictly needed, 
although it helped in removing the instrumentally caused curvature 
from the spectra.
The resulting light curves are depicted in Fig.\ \ref{fig:contlc}.
The error bars in this plot contain both the continuum
uncertainties due to photon shot noise and the error of the line flux 
rescaling factor. The zeropoint for both components 
is arbitrarily set at the continuum
magnitude of A in May 1993, which constitutes the brightest point.

Although the light curve is certainly not well-sampled, some features
are very cleary discernible. The strong decline in component A between
1993 and 1995, spanning almost a magnitude, is well mirrored in B,
except for the inflection between Nov 1994 and Feb 1995 which occurs
only in B. This feature alone is already very suggestive that B leads
the variability, as one expects from the observed lens configuration
(cf.\ R98). From early 1996 on, 
the sampling improved due to the beginning of regular monitoring,
and it became apparent that the object shows also significant
variability on relatively short time scales.

\begin{figure}
  \epsfig{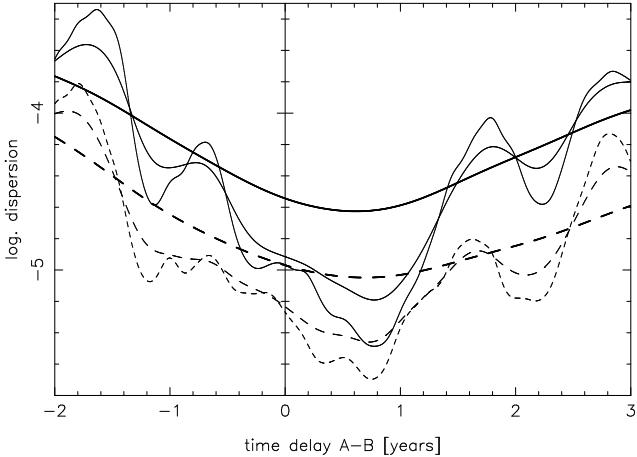}
  \caption[]{Dispersion curves for decorrelation lengths of 50 (thin),  
  100 (normal), and 365 (thick lines) days. 
  For the solid lines, the dispersion was
  calculated using statistical weights from the measurement errors.
  For the dashed lines, constant weights were used.}
  \label{fig:disp}
\end{figure}

\section{Time delay estimation}

Sliding the light curves against each other to find the best visual
agreement lead us to an estimate of about half a
year for the time delay A$-$B (component B leading) with a 
brightness difference of about 1.7 mag. For such a time delay, 
the observing periods of the one component coincide more or less 
with the seasonal gaps in the light curve of the other.
This effect without doubt produces a bias for values around
$\Delta t\approx 0.5\,\mathrm{yr}$, and we had to find a quantitative
method of estimation that is insensitive to such biases.

The dispersion method of Pelt et al. (\cite{pelt94}; \cite{pelt96}) 
proved to be quite robust despite the windowing effects in the case
of time delay determination for the double quasar 0957+561.
To minimise the bias for our data, we used the dispersion $D_{4,3}$
(Pelt et al. \cite{pelt96}) which takes into account not just
neighbouring data but uses the complete light curves with weight
factors corresponding to the (shifted) time differences of two
observations each. If the decorrelation length $\beta$ is chosen
sufficiently large, the windowing effect becomes increasingly less
significant. Figure~\ref{fig:disp} shows the dispersion for different
values of $\beta$ over an interval of \emph{a priori} possible time delays.

The global minima of all dispersion curves are consistently 
located between 0.6 and 0.8~years, which indicates a value of the
time delay in this range. Using a decorrelation length of 100~days 
which reduces the windowing but does not smooth the data on longer 
time scales, the minimum is at $\Delta t=0.73\,\mathrm{yr}$ 
with magnitude difference of 1.70 for the weighted data.
Note that the existence of slightly weaker local minima 
still permit a somewhat smaller time delay of 0.3--0.5~yrs. 
On the other hand, a $\Delta t$ of one year or more is not consistent
with the data.

This estimate of a time delay is to be seen as a very preliminary
result. A detailed analysis of the light curves has to wait until
better sampled data are available.

\begin{table}
  \caption{Relative positions of the images and the lensing galaxy,
           taken from the CASTLes compilation
           (Kochanek et al. \cite{castle}; 
            Leh\'ar et al., in preparation). 
            The directions of positive $x$ and $y$ are west and
            north, respectively. }
    \label{tab:pos}
    \begin{tabular}{lr@{.}lr@{.}l}\hline\noalign{\smallskip}
      & \multicolumn{2}{c}{$x\;[\arcsec]$} & 
        \multicolumn{2}{c}{$y\;[\arcsec]$} \\
      \noalign{\smallskip}\hline \noalign{\smallskip}
      A & 0&0 & 0&0 \\
      B & $-2$&$901\pm0.003$ & $-1$&$332\pm0.003$ \\
      G & $-0$&$974\pm0.003$ & $-0$&$510\pm0.003$ \\ \noalign{\smallskip}
      \hline
    \end{tabular}
\end{table}

\begin{table}
  \caption{Parameters for the SIST and the SIEMD model:
      $\alpha_0$ is the Einstein radius, $\epsilon$ and $\gamma$
      are ellipticity and shear, with position angles 
      $\theta$ and $\phi$, respectively.
      $M_{\mathrm{A}}$ and $M_{\mathrm{B}}$ are the magnifications,
      the signs giving the parity, and $T$ 
      contains the time delay according to Eq.\ \ref{eq:timedelay}.}
  \label{tab:modpar}
  \begin{tabular}{llr@{.}lr@{.}lll}\hline\noalign{\smallskip}
    model & $\alpha_0\;[\arcsec]$ & \multicolumn{2}{l}{$\epsilon$} & 
    \multicolumn{2}{l}{$\gamma$} &
    $M_\mathrm{A}$ & $T\;[\mathrm{arcsec}^2]$ \\
    & & \multicolumn{2}{l}{$\theta\;[\degr]$} & 
      \multicolumn{2}{l}{$\phi\;[\degr]$} &
    $M_\mathrm{B}$ & \\
\noalign{\smallskip}\hline\noalign{\smallskip}
SIST  & 1.403 & 0&    & 0&122 & $-7.23$ & 1.395 \\
      &       & \multicolumn{2}{c}{\ }& 112&1 & $+2.52$ & \\[1ex]
SIEMD & 1.546 & 0&191 & 0& & $-7.12$ & 1.590 \\
      &       &22&1   & \multicolumn{2}{c}{\ } & $+2.48$ & \\
      \noalign{\smallskip}\hline
  \end{tabular}
\end{table}

\section{Modeling the deflector potential}

The observational parameters of this system are the image and galaxy
positions (cf.\ table~\ref{tab:pos}), and the flux ratio of the images.
For the latter it is important to realise
that the continuum ratio of 4.8 (1.7\,mag) differs from the
emission line value because of the excess continuum component in image
A, tentatively identified with microlensing by Wisotzki et al.\
(\cite{wisotzki93}).
We use the flux ratio of 2.85 obtained from the emission lines (see above),
which should be much less affected by microlensing than the continuum
because of the larger size of the emitting region.

As a reference model, we use a singular isothermal sphere with
external shear (SIST). This is probably the simplest model capable to
reproduce the observed positions and the flux ratio (cf.\ R98). 
The parameters of this model can be found in
Table \ref{tab:modpar}. The observational uncertainties lead to an
internal error of only 0.6\,\% for the time delay.
To examine the much larger possible errors due to the modeling, 
we used a more general approach of models consisting of a singular isothermal
ellipsoidal mass distribution (SIEMD, see Kassiola \& Kovner
\cite{kassiola93}) with external shear. For the model-fitting we fixed
ellipticity $\epsilon$ and shear $\gamma$, 
and used the other parameters listed in
Table~\ref{tab:modpar} (including the position angles) and 
the source position to fit the observations. 
Due to the small number of constraints, the position of the lensing
galaxy was fixed at the observed values.
This was carried out for a range of
values for $\epsilon$ and $\gamma$. With the restriction of
$\epsilon<0.3$ and $\gamma<0.2$, we find a maximum deviation of
$+20\,\%$ and $-10\,\%$ for the time delay. As an example, the model 
with zero external shear is also given in Table~\ref{tab:modpar}. 
The time delay for this case is 14\,\% larger than for the SIST model.

\section{Implications for the lensing galaxy}

\begin{figure}
  \epsfig{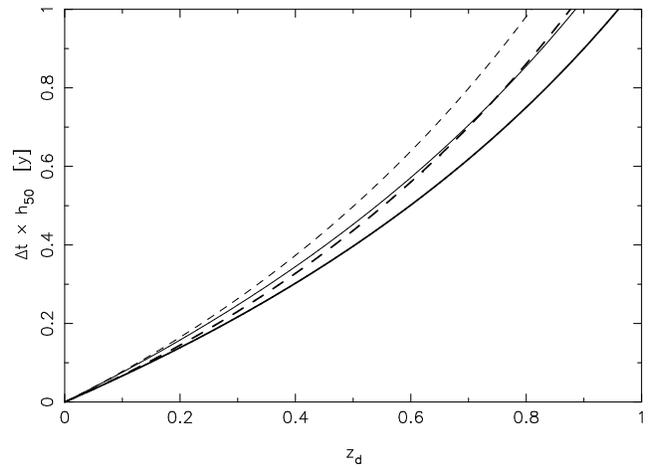}
  \caption[]{Time delay scaled by $H_0$ for the SIST (thick) and SIEMD (thin)
  model. Standard Einstein-de~Sitter cosmology is shown as solid
  lines, a low-density universe ($\Omega=0.3,\lambda=0$) as dashed
  lines. Flat low-density world models are located between these two curves.}
  \label{fig:timdelz}
\end{figure}

If the redshift of the lens $z_\mathrm{d}$ were known, we could compute
the Hubble parameter from the time delay using the fundamental equation
connecting the time delay $\Delta t$ and the parameter $T$
\begin{equation}
\Delta t \:=\: H_0^{-1}\,
(1+z_\mathrm{d})\,\frac{d_\mathrm{d}d_\mathrm{s}}{d_\mathrm{ds}}\,T \quad .
\label{eq:timedelay}
\end{equation}
As $z_\mathrm{d}$ is still unknown, we cannot use this formula to
get $H_0$. However, simply \emph{adopting} a canonical value for $H_0$ allows 
us to predict the redshift of the lens, 
or better, to constrain the range of possible values for $z_\mathrm{d}$.
Figure~\ref{fig:timdelz} shows the product of time delay and Hubble
parameter as a function of $z_\mathrm{d}$.
For $\Delta t=0.73\,\mathrm{yr}$,
$H_0=50\mathrm{\,km\,s^{-1}\, Mpc^{-1}}$, $\Omega=1$, and $\lambda=0$,
the SIST model predicts $z_\mathrm{d}=0.79$, and
the velocity dispersion of the galaxy for this model is 
$\sigma_v=332\mathrm{\,km\,s^{-1}}$. This corresponds to a mass of 
$9.2\,10^{11}\,M_\odot$ inside of one Einstein radius, well within 
the range expected for a reasonably massive galaxy. 
With an $I_c$ band magnitude of 20.9 according to R98, 
the mass-to-light ratio would then be of the order of 10 solar units,
again quite consistent with the expectations for such a galaxy
(cf.\ Keeton et al.\ \cite{keeton98}).

Recently, values for the time delay have been predicted based
on the assumption that one of the two strong metal absorption line 
systems at $z=1.32$ or $z=1.66$ can be identified with the deflector.
R98 give $\Delta t \simeq 1.9\,h_{50}^{-1}$~yrs,
Courbin et al. (\cite{courbin98}) even $3.5\,h_{50}^{-1}$~yrs.
Since we can reliably exclude $\Delta t > 1$\,yr,
our results are not compatible with $z_{\mathrm{d}}$ significantly
larger than 1; in particular, the absorbers at 1.32 and 1.66 
can be ruled out.

We have searched our higher resolution
NTT spectra of HE~1104$-$1805
(cf.\ Lopez et al.\ \cite{lopez98}) for absorption lines within the
redshift range permitted by Fig.\ \ref{fig:timdelz}. 
An additional demand is that the lines should be stronger in A, 
as this component is located closer to the deflector. 
Two Mg\,{\small II} absorption systems, at $z=0.52$ and 0.73, 
meet the criteria.
Of these, $z_\mathrm{d}=0.52$ is acceptable only for a time delay as 
short as $\sim 0.4$\,yrs, and is furthermore not compatible with 
the $I-K$ colour estimate of R98.
This leaves the system at $z=0.73$ as candidate; however, 
the lens could also be an elliptical galaxy for which
Mg\,{\small II} absorption would be a poor indicator. The very red
colours measured by R98 support such a notion.

\begin{acknowledgements}
We are grateful to all ESO staff and visiting astronomers
for their cooperation in obtaining the monitoring data. 
We also thank Dr.\ P.~Schechter for enlightening discussions 
and Dr.\ E.~Falco for providing us with the
latest astrometric data of the system.
\end{acknowledgements}


\begin{thebibliography}{}
\bibitem[1998]{courbin98} Courbin F., Lidman C., Magain P., 1998,
  A\&A 330, 57 
\bibitem[1993]{kassiola93} Kassiola A., Kovner I., 1993, ApJ 417, 450
\bibitem[1998]{keeton98} Keeton C.R., Kochanek C.S.., Falco E.E.,
  1998, ApJ in press, astro-ph/9708161
\bibitem[1998]{castle} Kochanek C.S., Falco E.E., Impey C., et al.,
  1998, CASTLe Survey, {\tt<http://cfa-www.harvard.edu/castles/>}
\bibitem[1998]{lopez98}
  Lopez S., Reimers D., Rauch M., Sargent W.L.W., Smette A.,
  1998, ApJ in press, astro-ph/9806143
\bibitem[1994]{pelt94} Pelt J., Hoff W., Kayser R., Refsdal S., et
  al., 1994, A\&A 286, 775
\bibitem[1996]{pelt96} Pelt J., Kayser R., Refsdal S., Schramm T.,
  1996, A\&A 305, 97
\bibitem[1998]{remy98} Remy M., Claeskens J.-F., Surdej J., Hjorth
  J., et al., 1998, New Astronomy 3, 379 
\bibitem[1995]{smette*95} Smette A., 1995, in: QSO Absorption Lines,
ed.\ G. Meylan, ESO Astrophysics Symposia, 275
\bibitem[1993]{wisotzki93} Wisotzki L., K\"ohler T., Kayser R.,
  Reimers D., 1993, A\&A 278, L15
\bibitem[1995]{wisotzki95} Wisotzki L., K\"ohler T., Ikonomou M.,
  Reimers D., 1995, A\&A 297, L59
\end{thebibliography}
\end{document}